# FHiVI²: A Fast High–Voltage approach for Ionisation Improvement in the context of plasma–assisted magnetron sputtering

R. JEAN–MARIE–DESIREE[1], L. De POUCQUES[1], S. CUYNET[1]

[1] *Institut de Jean Lamour, Université de Nancy, Nancy, France*

**Abstract:** Ionisation control remains a key challenge in plasma–assisted processes and more particularly in magnetron sputtering. In this context, the former directly governs species transport and thin–film properties. Although HiPIMS substantially enhances the ionisation rate of sputtered species, its performance remains constrained by the coupling between sputtering and ionisation, which limits discharge tunability. Here, we introduce Fast High Voltage for Ionisation Improvement (FHiVI²), a new transient electrical regime designed to improve the ionisation rate without requiring additional elements. By applying an *ad hoc* very short, high–voltage pulse during plasma sputtering processes, FHiVI² regime is expected to induce a strongly anisotropic electric–field burst intended to enhance sputtered metallic species ionisation while limiting back–attraction losses and sputtering. To assess the ionising potential of this regime, optical emission spectroscopy together with mass spectrometry were carried out. More particularly, mass spectrometry measurements show an improvement of the ionisation rate by a factor of 2.7 compared with the HiPIMS regime, for an applied voltage of 2 kV which correspond to an additional mean power of only 1.2 W, using a 2–inch tungsten target. These findings suggest that FHiVI² provides a flexible route towards improved ionisation control in magnetron sputtering and opens new perspectives for plasma–assisted processes.

## 1. Introduction

In the study of ionised media, and more specifically non–equilibrium plasmas, the concept of ionisation rate, as well as the selectivity of ionisation processes in complex gas mixture, is of major importance in the context of their applications. Indeed, the nature of the species produced, their energy distribution, and their density determine many functional and manufacturing processes. Several examples may be cited, including ion propulsion [1, 2], plasma etching processes [3, 4], semiconductor doping processes [5–7], and the design of thin films exhibiting specific mechanical, electrical and/or optical properties [8, 9]. In this context, a thorough understanding and control of ionisation mechanisms are key parameters to optimising the performance of plasma–assisted processes.

Within this field, the plasma–assisted deposition community has developed numerous strategies for enhancing the ionisation rate of sputtered species, thereby enabling ultimately fine control the incident species flux and the desired film properties. Known as IPVD (ionised physical vapour deposition) [10, 11], some of these processes rely on optimising the probability of electron–(metal) neutral collisions by extending the electron residence time through magnetic confinement or by increasing electron density. In the literature, the latter approach has been realised by implementing secondary plasma sources, either radio–frequency or microwave, within the sputtering cone [12–15]. However, this type of technology requires the introduction of additional elements into the reactor, such as coils, magnets, or waveguides, which can induce modifications that are sometimes substantial. Moreover, these components are themselves subject to sputtering, thereby constituting a potential source of contamination and rendering the approach less adaptable.

A second strategy for increasing electron density consists in applying a high–power density over relatively short pulse durations $T_{\mathrm{HiPIMS}}$, typically of several tens of microseconds, at repetition frequencies of up to a few kilohertz [16]. This defines the HiPIMS regime, whose operation does not require the insertion of additional elements into the reactor. Compared with other strategies, this regime is simpler to implement and can yield ionisation fractions that unfortunately vary considerably depending on the experimental conditions (target, gas, etc.) [17–20].

In practice, the HiPIMS regime offers only limited control over the ionisation fraction of the sputtered vapour, which in turn governs the properties of the growing film. Several studies have reported a correlation between pulse duration $T_{\mathrm{HiPIMS}}$ and the resulting microstructure [21–23]. Because pulse duration, frequency, and duty cycle affect both the ionised flux and the sputtering dynamics, the optimisation of ionisation remains intrinsically constrained.

More recently, J. Hnilica and co–workers [24–26] proposed a burst–mode operation, referred to as multi–HiPIMS, in which a conventional HiPIMS pulse is followed by a train of similar HiPIMS pulses. This strategy introduces two characteristic time scales: a macroscale related to the overall burst period and a microscale one related to the spacing between pulses. This approach seeks to exploit the recovery time of the former discharge to maintain favourable ionisation conditions while improving deposition efficiency. Crucially, their innovative work brings to the fore the sputtered–species transport and the time window in which ionisation occurs. Nevertheless, this mode of operation remains fundamentally coupled to the sputtering process and therefore provides only limited control over the ionisation rate of the sputtered metallic species. Indeed, the close coupling between sputtering and ionisation makes it more difficult to achieve high ionisation rates. As the density of sputtered metal species increases, the electron temperature may be reduced, thereby decreasing the ionisation efficiency. Further efforts have been devoted to coupling HiPIMS with a so–called pre-ioniser unit. This unit typically consists of either a direct–current background regime [27–29] or a pulsed direct–current background regime (referred to as low–power pulse by N. Brenning *et al.* [30]) background that provides seed charges as well as sputtered species to be ionised. However, the constant negative voltage applied to the cathode does not necessarily eliminate back-attraction, which remains a key limitation for ionisation yield.

Thus, in the context of improving the ionisation rate in magnetron sputtering, developing a strategy based on an electrical regime that does not rely on additional reactor components appears necessary. In addition and considering the foregoing state–of–the–art discussion, such a regime should ideally satisfy at least two phenomenological criteria to enable efficient ionisation: a negligible sputtering yield and limited back-attraction of the produced metallic ions. These two criteria led us to design a highly transient pulsed regime with a very high–voltage amplitude, named here as Fast High Voltage for Ionisation Improvement (FHiVI²). By combining a very short pulse duration and a very fast rise time with a very high voltage amplitude at the cathode, this approach is expected to generate a rapid, strongly anisotropic electric–field burst, enabling electron acceleration in the target vicinity. The resulting electric–field variation is anticipated to reach amplitudes several times, or even several orders of magnitude, greater than those associated with the regimes discussed above. Therefore, the induced electron–(metal) neutral ionisation collisions occur on a timescale shorter than that of the ion plasma frequency $\omega_{\mathrm{pi}}$. In other words, this regime must therefore allow ions to be produced in the volume while limiting back–attraction losses and sputtering processes. One may note that this new FHiVI² regime has been designed to offer adaptability and flexibility over multiple plasma processes, as the FHiVI² module can be implemented across a wide range of existing regime and plasma–assisted processes (DC, RF, and even DBD for atmospheric processes).

In the present study, the FHiVI² electrical module has been designed and adapted for operation with a conventional HiPIMS system. Since HiPIMS is arguably the most widely used approach for addressing the challenges associated with IPVD, this regime is taken here as the reference against which the beneficial effects of the additional FHiVI² regime are assessed solely in terms of ionisation.

## 2. Experimental setup

The experimental setup section briefly describes the sputtering reactor assembled for this study, the operating system and the synchronisation scheme of the different regimes used, and finally the different diagnostic tools developed and employed to investigate them.

The chamber consists of a 6 × $10^{-2}$ $m^3$ grounded stainless–steel chamber. The atmosphere inside is pumped down to a pressure of about $10^{-4}$ Pa using a dedicated pumping system. Once this vacuum is reached, the chamber is filled with argon gas (Air Liquide ALPHAGAZ 2, over 6N purity grade) up to the working pressure set to 1 Pa. A guillotine valve isolating the pumping system from the chamber allows fine regulation of the working pressure which is monitored using a capacitive gauge (Pfeiffer CMR 375). The high–voltage is applied through a water–cooled circular magnetron device (Gencoa 3G Circular) on which is installed a 3–mm–thick, 2–inch–diameter tungsten target (Neyco W2PX3MM3N5, ~99.95% of purity) to sputter.

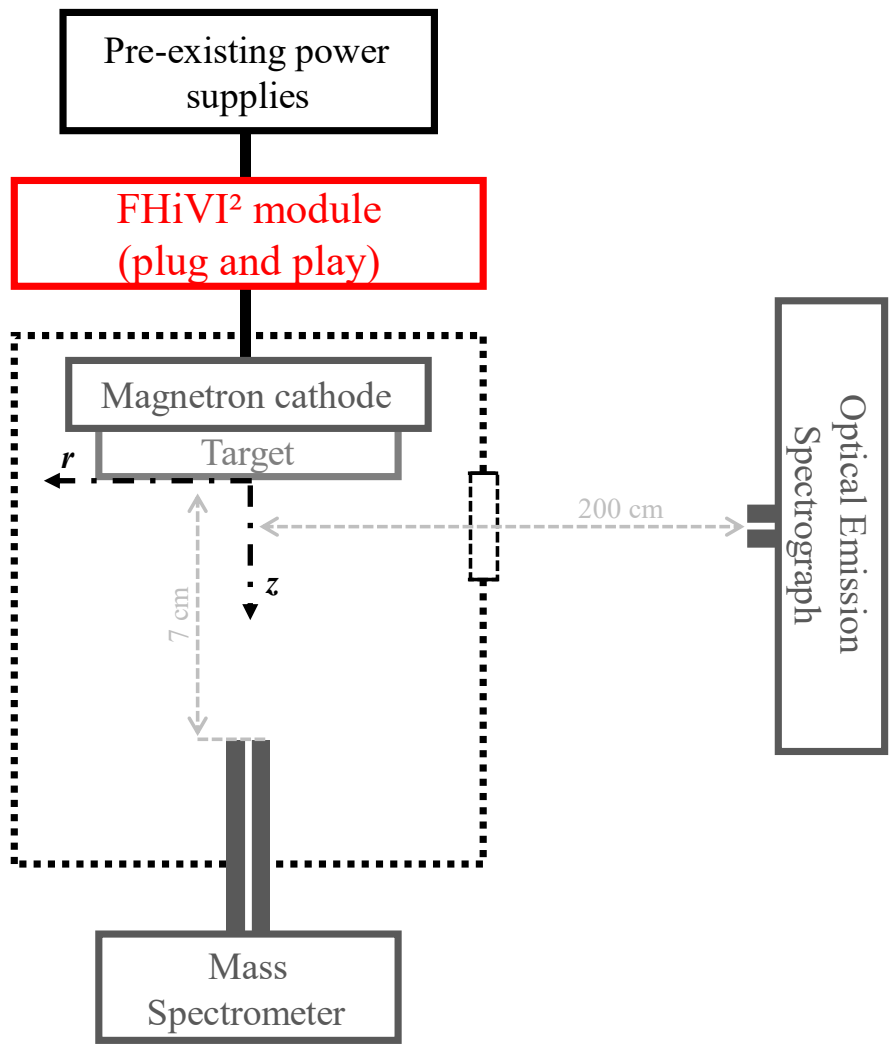

*Figure 1: Diagram showing the integration of the FHiVI² module into an existing magnetron sputtering system, together with the main diagnostic techniques presented in this study: mass spectrometry and optical emission spectroscopy.*

In this study, three types of pulsed electrical regimes were implemented to operate together or individually: HiPIMS, FHiVI² and PP (Positive Polarisation). A general timing diagram showing the synchronisation of these three regimes operating together is presented in Figure 2.

The HiPIMS regime is powered by a MELEC SIPP 2000 high–voltage power supply, capable of delivering a maximum voltage of 1 kV at 5 A direct current. To shape the voltage waveform, the power supply is driven by an external arbitrary function generator (Tektronix AFG31022), designated as the master. In this study, the resulting negative pulse lasts $T_{\mathrm{HiPIMS}}$ = 20 $\mu$s with a repetition frequency $\nu$ = 200 Hz, corresponding to a period $T$ = 5 ms, as described in more detail in section 3.1.1.

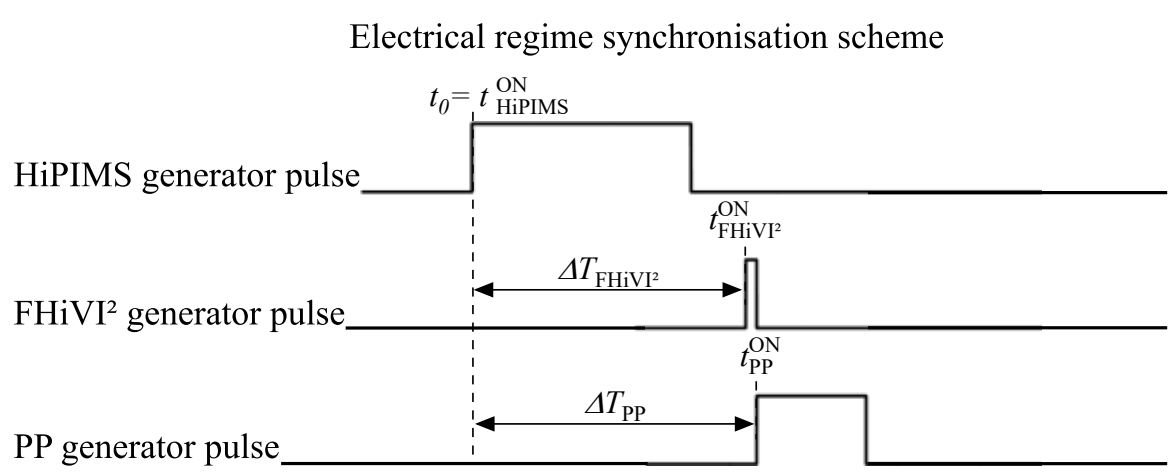

*Figure 2: Timing diagram of control signals for the studied pulsed regimes (HiPIMS, FHiVI², and PP).*

The FHiVI² regime is powered by an independent direct high–voltage generator (Technix SR15–R–1200) capable of delivering a maximum negative voltage of 15 kV at a maximum power 1.2 kW. Within the FHiVI² module, this negative DC voltage is chopped by a high–voltage, high–frequency switch (Behlke HTS 301–03–GSM). This FHiVI² power unit is also controlled by an external arbitrary function generator (Tektronix AFG3022C), which is itself phase–locked and synchronised with the one driving the HiPIMS generator. The resulting negative pulse lasts $T_{FHiVI^2} = 1\ \mu$s with a repetition frequency $\nu$ set by the master AFG. Although this FHiVI² regime can be triggered at any user-defined time, either before, during, or after the HiPIMS pulse, regardless of the sputtering regime, it is voluntarily triggered 14 $\mu$s after the HiPIMS off-time, allowing the two negative pulses to be well isolated from each other, as described in more detail in section 3.1.3. For the purposes of this study, and due to certain limitations on the measuring instruments used, the FHiVI² voltage is intentionally limited to 2.5 kV. However, the FHiVI² module can still reach the maximum voltage given by the high–voltage generator (- 15 kV).

As for the PP, this regime is powered by an independent regulated–voltage generator (Elektro–Automatik PS 5200–10 A) out of the HiPIMS generator and capable of delivering a maximum positive voltage of up to 0.2 kV at a maximum power 0.64 kW. The FHiVI² module schematised in Figure 1 also controls this positive DC voltage which is likewise chopped by a high–voltage, high–frequency switch (Behlke HTS 301–03–GSM). The resulting positive pulse lasts $T_{PP}$ = 10 $\mu$s at $\nu$, triggered after all negative pulses have been applied, when both the voltage and current measured during the HiPIMS regime are zero and corresponding to the very end of the FHiVI² regime, as described in more detail in section 3.1.2. Obviously, its application is analogous to that of bipolar HiPIMS effect.

The combination of these different regimes results in the parallel connection of the switches for the HiPIMS and FHiVI² regimes, themselves in series with that of the PP regime and, of course, the cathode. However, those electrical associations cause a drop in output potential (voltage divider effect) due to the non–negligible intrinsic impedance of the system. In addition, paralleling high–voltage generators require the implementation of resistive elements dimensioned to handle such power levels. Therefore, the implementation of these different pulsed regimes requires several precautions, both in terms of the electrical circuit design and in the temporal management of the various pulsed power supplies formed by the generator–switch pairs described above. Ultimately, the power supply developed through in–house switch integration, combined with a carefully selected set of power electronic components, enables the generation and modulation of the three distinct pulsed regimes. Thus, all these regimes share a common reference point $t_0$, corresponding to the onset of the HiPIMS pulse ($t_{\mathrm{HiPIMS}}^{\mathrm{ON}}$), as illustrated in Figure 2. The notations $\Delta T_{\mathrm{FHiVI^2}}$ and $\Delta T_{\mathrm{PP}}$ respectively denote the delay in applying the FHiVI² ($t_{\mathrm{FHiVI^2}}^{\mathrm{ON}}$) and PP ($t_{\mathrm{PP}}^{\mathrm{ON}}$) regimes relative to $t_{\mathrm{HiPIMS}}^{\mathrm{ON}}$. Our configuration provides full control over these delays, offering an additional degree of freedom for investigating the influence of the FHiVI² regime and its interaction with the others.

The study of the influence of the FHiVI² regime and its effectiveness in terms of ionising capacity required the implementation and development of several time–resolved diagnostics.

One of the most common diagnostics is the measurement of the voltage–current characteristics, carried out at the cathode using a CalTest Electronics CT4028 voltage probe (bandwidth: DC – 220 MHz) and a MagneLab CT1.0 current probe (bandwidth: 200 Hz – 500 MHz). Both signals were recorded on a LeCroy WaveSurfer 4104HD digital oscilloscope (bandwidth up to 1 GHz), averaged over 20 sweeps in this study.

Along with these electrical measurements, tungsten emission lines were also analysed using a Jobin–Yvon Triax 550 optical spectrometer equipped with a 1800 gr.mm$^{-1}$ grating and an i-Spectrum Two iCCD camera. Since the objective is to assess the influence of the pulsed FHiVI² regime on the sputtered tungsten species in time, the measurements were time–resolved. To achieve this, the iCCD detector was triggered by the second channel of the master AFG, ensuring synchronisation with the discharge. Given that the FHiVI² regime lasts less than 1 $\mu$s, the detector integration time was set to 0.2 $\mu$s, providing a comparable temporal resolution. Consequently, the measurements were performed over a time range of 400 $\mu$s with a step of 0.2 $\mu$s for all the studied FHiVI² operating conditions.

To investigate the tungsten $W^+$ ion fluxes generated by these regimes, and more specifically by the FHiVI² regime, a Hiden EQP–20 300 mass spectrometer was installed on the reactor to perform time–resolved $W^+$ ion energy distribution analyses. The grounded orifice of the spectrometer, 100 $\mu$m in diameter, was aligned with the cathode axis and positioned 70 mm away from the target surface. In this position, the mass spectrometer is sufficiently far from the plasma source to minimise disturbances particularly due to the FHiVI² regime application, while its orifice remains exposed to the influence of the entire erosion track. One should note that adjusting the voltage settings of the various components of the mass spectrometer—particularly the "EQP extractor" grid positioned just behind the orifice—requires the utmost care, as these settings strongly influences the energy distribution profile of the ions passing through the instrument. To ensure time–resolved measurements, the secondary electron multiplier detector of the spectrometer was triggered by the master AFG, guaranteeing synchronisation with the discharge. The detector integration time was set to its minimal value of 1 µs, which does not fall below $T_{\mathrm{FHiVI^2}}$. This temporal resolution is nevertheless sufficient to characterise the $W^+$ ion flux entering the mass spectrometer, whose transport has been significantly altered since its generation in the plasma. In such a configuration and considering the transit time (or time of flight) of $W^+$ ions inside the spectrometer, the timescale corresponds only to the arrival of the $W^+$ species at the orifice upon $t_{\mathrm{HiPIMS}}^{\mathrm{ON}}$. In all cases, the measurements result from the accumulation of signals over 20 acquisitions, averaged three times for a total of 60 consecutive discharges.

Finally, the FHiVI² regime was used for the very first time in combination with the HiPIMS and the PP regimes under specific conditions discussed in section 3.1.3. The aim of this combination is to assess the feasibility and relevance of the FHiVI² regime in terms of ionisation efficiency, with reference to HiPIMS as the most studied and effective pre–existing regime.

## 3. Results and discussion

To begin with, the discussion focuses on the voltage–current characteristics of the three different regimes used and progressively combined: HiPIMS, FHiVI² and PP, which can be described respectively through time–domains *I*, *II* and *III*. Then, the results from optical emission spectroscopy (OES) and mass spectrometry (MS) are presented and discussed to evaluate how efficiently the FHiVI² regime ionises the sputtered metallic species.

### 3.1. Voltage–current (V–I) characterisation

The V–I characteristics of the three regimes are investigated and discussed in detail in the following subsections. To highlight the novelty of FHiVI² characteristics, the HiPIMS regime was first operated alone to establish the baseline for our study, before being coupled with PP regime. The FHiVI² regime is then combined with both preceding regimes.

#### *3.1.1. HiPIMS–only regime*

As mentioned earlier, the HiPIMS regime serves two main purposes in this study. First, it provides a reference point, since the magnetron sputtering community recognises this regime for its high ionisation rate under suitable operating conditions. Second, it supplies the FHiVI² regime with sputtered metallic species as the former FHiVI² is intended to solely improve the ionisation rate here. In this section, the discussion is focused only on the HiPIMS regime alone.

Figure 3 shows the V–I characteristics measured at the cathode during a typical HiPIMS pulse, as configured for our study.

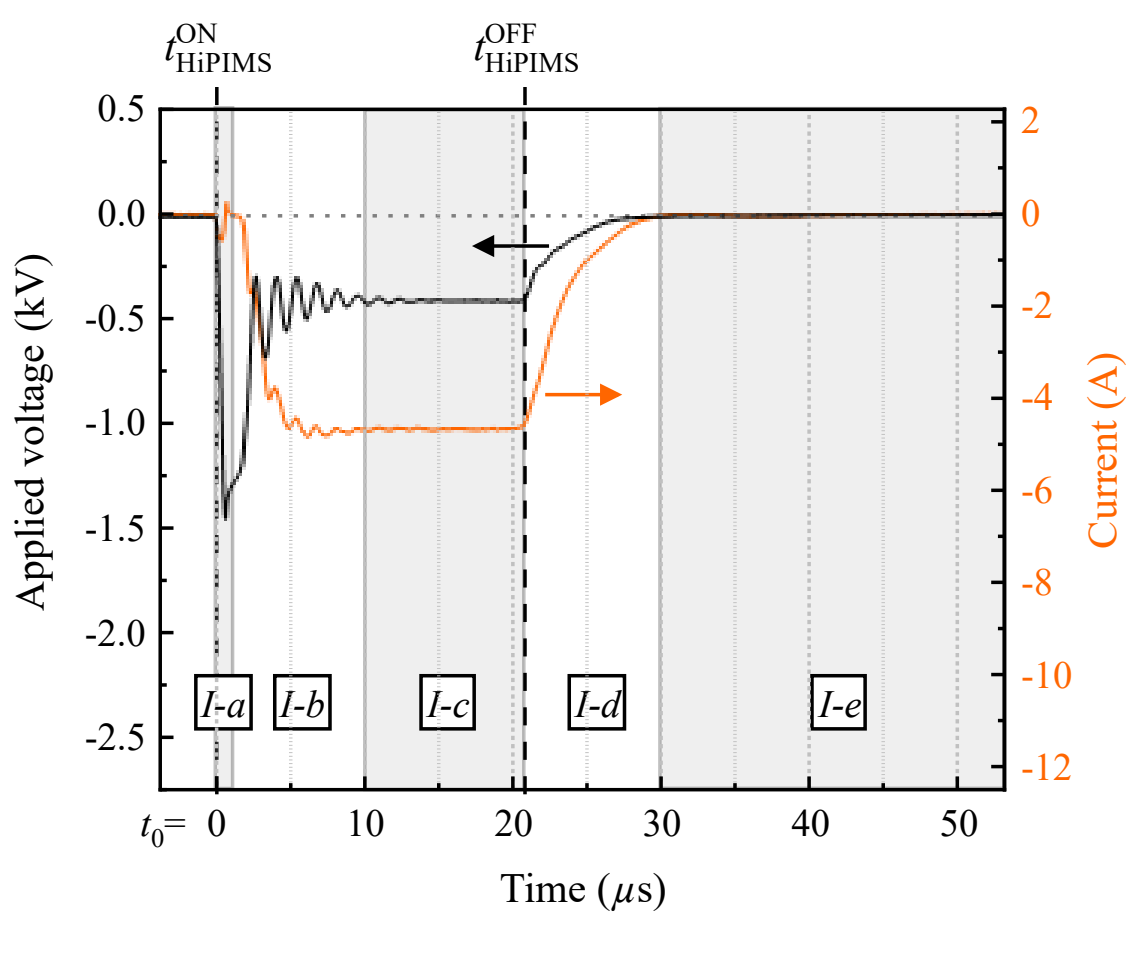

*Figure 3: Time evolution of the applied voltage (black) and the current (orange) for the HiPIMS–only regime over time–domain I, measured at the cathode. Measurements were performed in pure argon at 1 Pa with ν = 200 Hz, $T_{HiPIMS}$ = 20 µs and $V_{HiPIMS}$ = -0.4 kV.*

From $t_0$ = 0 $\mu$s, these V–I profiles could be distinguished in five different time–domains.

Time–domain *I-a* corresponds to a relatively intense and brief variation of the applied voltage. In our case, this corresponds to a voltage peak of approximately -1.3 kV reached within 1 $\mu$s. This voltage variation induces a small displacement current, as described by Maxwell–Ampère's law.

In the following time–domain *I-b*, the actual discharge establishes itself as indicated by the increase in the conduction current amplitude. This increase occurs concomitantly with a decrease in the applied voltage, which stabilises around -0.4 kV with periodic damping characteristics of displacement–current rebounds arising from reactance of the HiPIMS power supply [23, 31]. Additionally, the discharge current reaches a plateau value of -4.7 A from $t_0$ + 10 $\mu$s, marking the onset of the next time-domain *I-c*.

This time–domain is characterised by stable voltage and current values of -0.4 kV and -4.7 A, respectively, sustained for 11 $\mu$s. One of the commonly used criteria for characterising an HiPIMS regime is based on the maximum cathode current density, computed with the peak current and the active cathode surface, commonly referred to as the racetrack, where sputtering occurs. A current density reaching 1 $A.cm^{-2}$ is generally considered characteristic of HiPIMS [32]. To give some value, based on the current measurement shown in Figure 3 and an estimated active surface area of approximately 5 cm² for the 2-inch target, the maximum current density reaches in our case 0.94 $A.cm^{-2}$, which is close to this commonly used threshold.

Time–domain *I-d* describes the HiPIMS regime turn-off starting at $t_0$ + 21 $\mu$s. Both voltage and current decrease progressively in amplitude, approaching zero values following a typical capacitor discharge profile.

Finally, time–domain *I-e* corresponds to the period during which the voltage and current measured at the cathode are essentially null and invariant. For the HiPIMS–only regime, this time–domain extends until the next HiPIMS pulse at $t^{ON}_{HiPIMS} + T$, which corresponds to the beginning of a new time–domain *I-a*.

#### *3.1.2. HiPIMS and PP regime coupling*

The application of a positive polarisation (PP) pulse following the HiPIMS regime is analogous to the bipolar HiPIMS configuration. In our case, the aim of this coupling is to enhance the detection of plasma ions by repelling them away from the cathode surface and provide information on the volume charge distribution [33].

Figure 4a shows the V–I characteristics corresponding to the HiPIMS and PP regimes, measured at the cathode. As a reminder, PP is designed such that its timing can be entirely arbitrary from the technical perspective. In the present case, the PP regime is applied at $\Delta T_{PP} = t^{ON}_{HiPIMS} + 36$ $\mu$s, equivalent to $t^{ON}_{PP} = t_0 + 36$ $\mu$s, ensuring that the positive pulse does not overlap with the *I-(a-d)* time–domains associated with the HiPIMS pulse and better distinguish their respective effects. With $T_{PP}$ = 10 $\mu$s and a set voltage $V_{PP}$ = +150 V, the PP regime is therefore implemented within the *I-e* time–domain previously described. In this configuration, the measurements obtained during the PP regime mainly reflect the volume charge composition of the HiPIMS post–discharge region near the target surface.

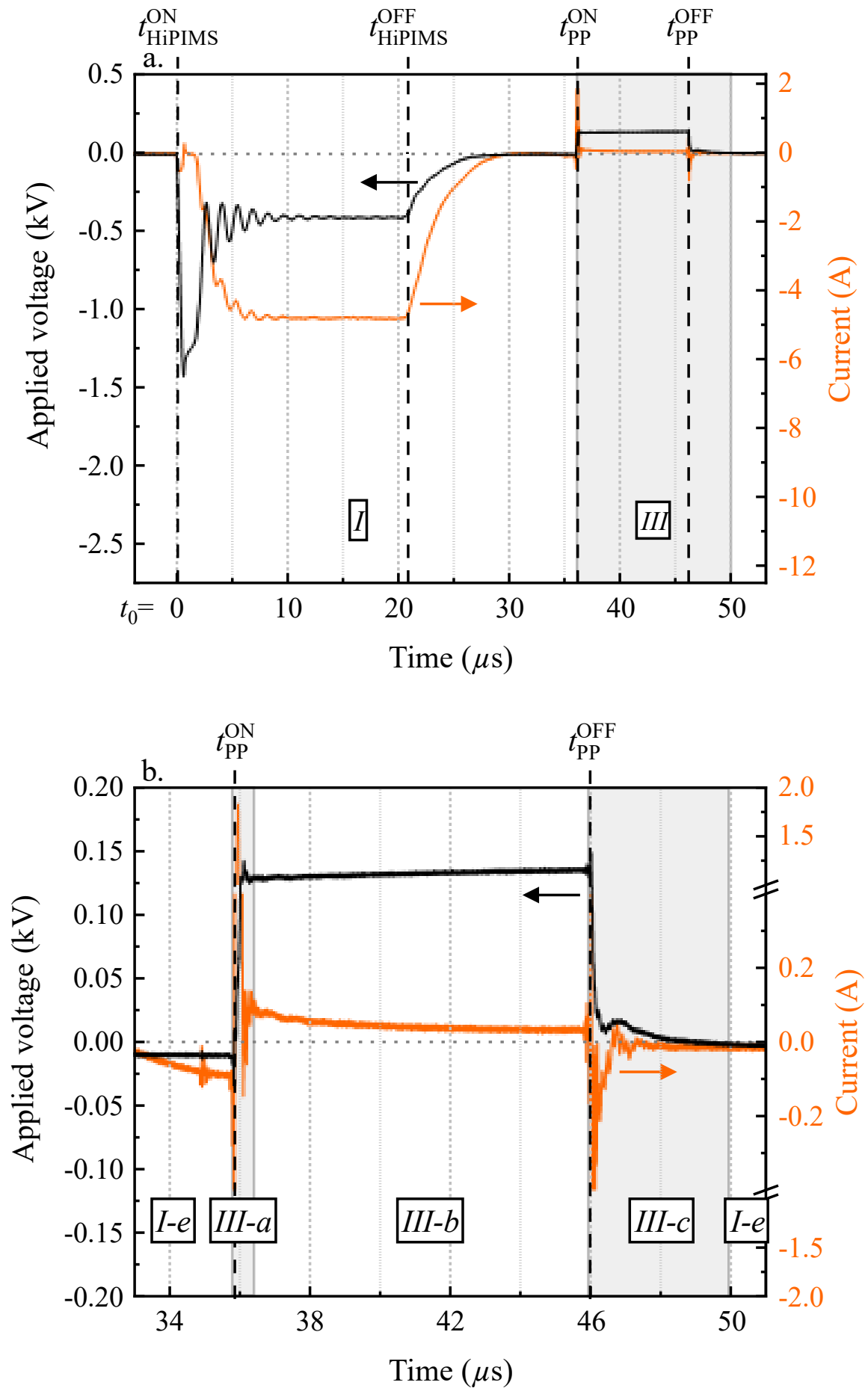

*Figure 4: Time evolution of the applied voltage (black) and the current (orange) for the combined HiPIMS + PP regime, measured at the cathode. (b) Close–up view around the PP regime to better discern its specific variations over time–domain III, with two current axis breaks. Measurements were performed in pure argon at 1 Pa with ν = 200 Hz, $T_{HiPIMS}$ = 20 µs, $V_{HiPIMS}$ = -0.4 kV, $T_{PP}$ = 10 µs, $V_{PP}$ = +150 V and $\Delta T_{PP}$ = 36 µs.*

Moreover, the addition of PP does not seem to have a significant influence on the V–I characteristics of the HiPIMS discharge. Indeed, since the voltage and current measured are already close to zero at the end of time–domain *I-e* (before each $t_{\mathrm{HiPIMS}}^{\mathrm{ON}}$), the subsequent HiPIMS discharge appears unaffected, as can be observed by comparing Figure 3 and Figure 4a.

The time–domain corresponding to PP regime is arbitrarily denoted as *III-*. Figure 4b provides a close-up view to better distinguish the different time-domains during PP regime, labelled from *III-a* to *III-c*.

The time–domain *III-a* begins with a sharp voltage rise to approximately +125 V within a few tens of nanoseconds. This abrupt increase occurs simultaneously with relatively large current variations (from -0.4 A to +2 A) up to around $t_0$ + 36.5 $\mu$s. Thus, time–domain *III-a* represents a transient regime, characterised by a brief and intense current primarily associated with a displacement current. This current is significantly higher than that previously observed during the HiPIMS regime (time–domain *I-a*): the temporal variation $\frac{dU}{dt}$, and consequently that of the electric field $\frac{dE}{dt}$, is much greater in domain *III-a*.

The time–domain *III-b*, during which the positive–polarisation current is measured, provides partial insight into the HiPIMS post–discharge in the vicinity of the target surface [33]. The measured current corresponds to a conduction current that decreases from +0.07 to +0.02 A, while the voltage increases slightly to reach approximately +135 V.

Time–domain *III-c* begins at $t_0$ + 46 $\mu$s, corresponding to the PP regime turn–off, *i.e.* $t_{\mathrm{PP}}^{\mathrm{OFF}}$. The applied voltage then rapidly drops before tending towards zero at approximately $t_0$ + 50 $\mu$s. Simultaneously, the measured current increases sharply in absolute value and is once again predominantly associated with a displacement current. Finally, when both voltage and current approach zero, their characteristics merge into time–domain *I-e*.

### *3.1.3. Combination of the three regimes: HiPIMS + FHiVI² + PP*

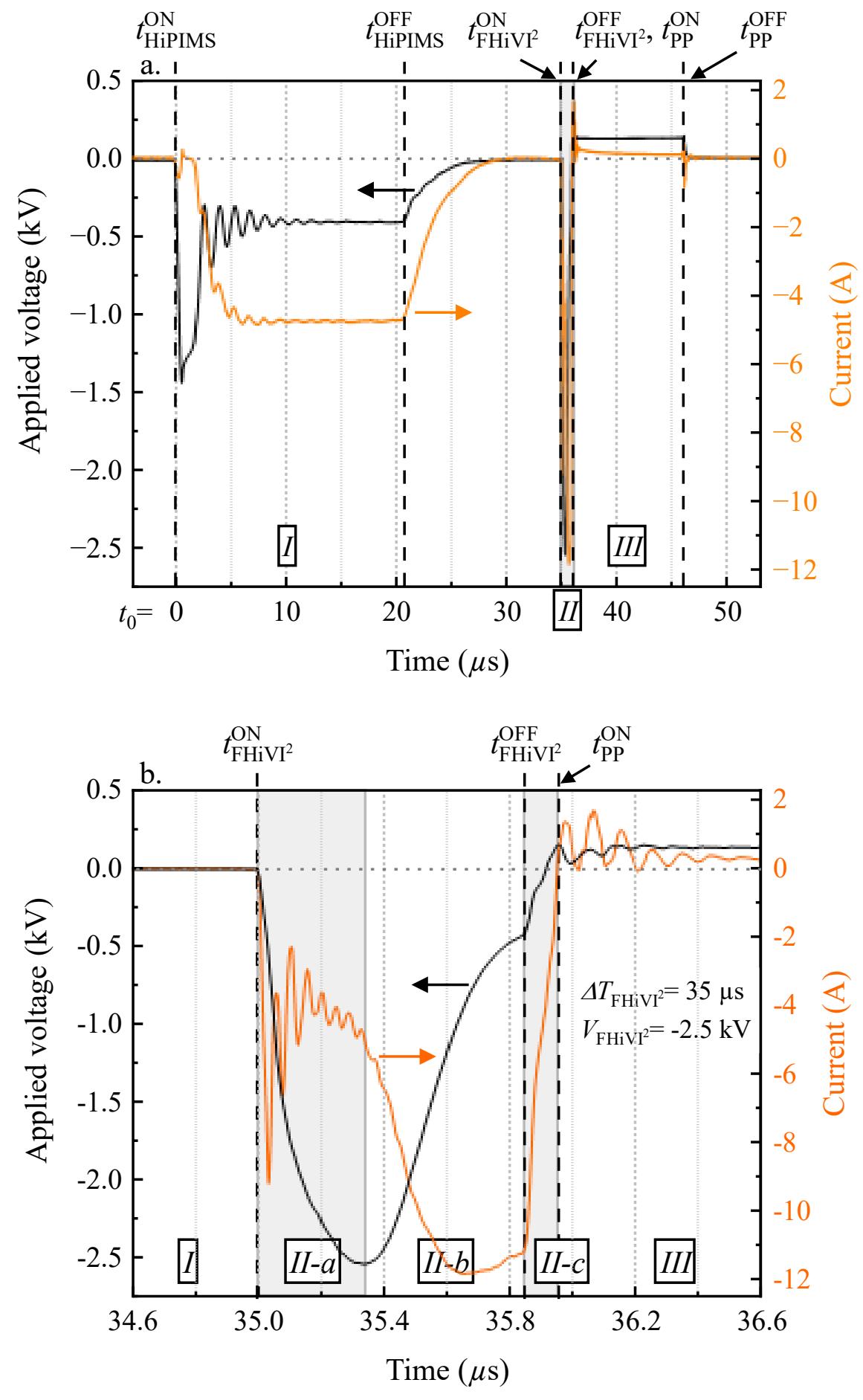

*Figure 5: Time evolution of the applied voltage (black) and the current (orange) for the combined HiPIMS + FHiVI² + PP regime. (b) Close–up view around the FHiVI² regime highlighting its temporal variations over time–domain II. Measurements were performed in pure argon at 1 Pa with $\Delta T_{FHiVI^2}$ = 35 µs and $V_{FHiVI^2}$ = -2.5 kV at ν = 200 Hz. The HiPIMS and PP regimes were operated under the following conditions: $T_{HiPIMS}$ = 20 µs, $V_{HiPIMS}$ = -0.4 kV, $T_{PP}$ = 10 µs, $V_{PP}$ = +150 V and $\Delta T_{PP}$ = 36 µs.*

As implemented, the FHiVI² regime is temporally independent and can therefore be applied at will. In this outset study, this on–demand high–voltage regime is used to increase the ionisation rate of the metal vapour produced by the HiPIMS sputtering regime. To this end, the FHiVI² regime is always applied between the HiPIMS regime (source of sputtered metal vapour) and immediately prior to the PP regime (used to probe the generated ions).

Following this scheme, Figure 5a shows a FHiVI² pulse timed at $\Delta T_{FHiVI^2} = t^{ON}_{HiPIMS} + 35\ \mu s = t^{ON}_{FHiVI^2}$, which corresponds to $t^{OFF}_{HiPIMS} + 14\ \mu s$. Such a delay in $t^{ON}_{FHiVI^2}$ allows the respective contributions of the HiPIMS and FHiVI² regimes to be distinguished. The time–domain describing the voltage (black) and current (orange) waveforms of the FHiVI² regime is then denoted as *II*. Figure 5b presents a close–up view around the FHiVI² regime highlighting its temporal variations. These variations can be distinguished into three time–domains, labelled *II-(a–c)*.

Time–domain *II-a* marks the onset of the FHiVI² regime from $t^{ON}_{FHiVI^2}$. This domain is characterised by a rapid increase in the applied voltage amplitude to reach the desired voltage $V_{FHiVI^2}$ = -2.5 kV in less than 340 ns. This results in strong current oscillations measured at the cathode that are associated with displacement–current rebounds arising only from the reactance of the FHiVI² circuitry. However, these oscillations damp rapidly within 340 ns settle at approximately -4 A at the end of *II-a*. As discussed for time–domain *III-b*, this -4 A value could be attributed to a conduction current arising from thermalised electrons produced earlier by the HiPIMS regime and subsequently strongly re-accelerated by the FHiVI² regime during *II-a*. With close look into the electrical characteristic reported in the literature, whenever an HiPIMS is combined with another electrical regime (*e.g.* pre-ionised HiPIMS, multi-HiPIMS) this result can also be observed in a very short timescale, upon the voltage rise [30].

Time–domain *II-b* begins at $t_0$ + 35.3 $\mu$s, corresponding approximately to the FHiVI² voltage maximum value reached ($V_{FHiVI^2}$) and the end of current damping. During this time–domain, a sharp increase in the current amplitude occurs to approximately 11.8 A within 500 ns. The maximum current density over the racetrack reaches 2.4 A.cm$^{-2}$ during the FHiVI² regime, despite operating at only 15% of the maximum available voltage (15 kV). This current density value clearly exceeds the commonly used criteria of 1 A.cm$^{-2}$ for characterising an HiPIMS regime [32]. Simultaneously, the applied voltage rapidly drops in amplitude by about ~2.1 kV from its peak value to -0.43 kV by the end of time–domain *II-b*. Once again, this voltage drop may be attributed to the power limitation of the high-voltage supply driving the FHiVI² regime. Consequently, the current decreases slightly by about 0.6 A after reaching its maximum value. However, the significant increase in current observed in time–domain *II-b* demonstrates that the FHiVI² regime is both rapid and highly effective for instantaneous charge production.

Finally, time–domain *II-c* starts around $t_0$ + 35.9 $\mu$s, corresponding to the turn–off of the FHiVI² regime $t^{OFF}_{FHiVI^2}$. From $t^{OFF}_{FHiVI^2}$, both voltage and current rapidly decrease in absolute value, approaching zero within slightly over 100 ns. The end of time–domain *II-c* thus marks the onset of the PP regime.

In summary, Figure 5 shows voltage rise and fall times on the order of a few hundred nanoseconds over a total application duration of ~1 $\mu$s during time–domain *II*, together with the rapid generation of a large conduction current. These V–I characteristics thus fully describe a typical operating FHiVI² regime, consistent with the intended electrical behaviour. Assessing the effect of the FHiVI² regime on the PP regime—and specifically on the positive polarisation current $I_{PP}$—provides relevant information regarding the modification of the charge density in the post–discharge, particularly near the cathode.

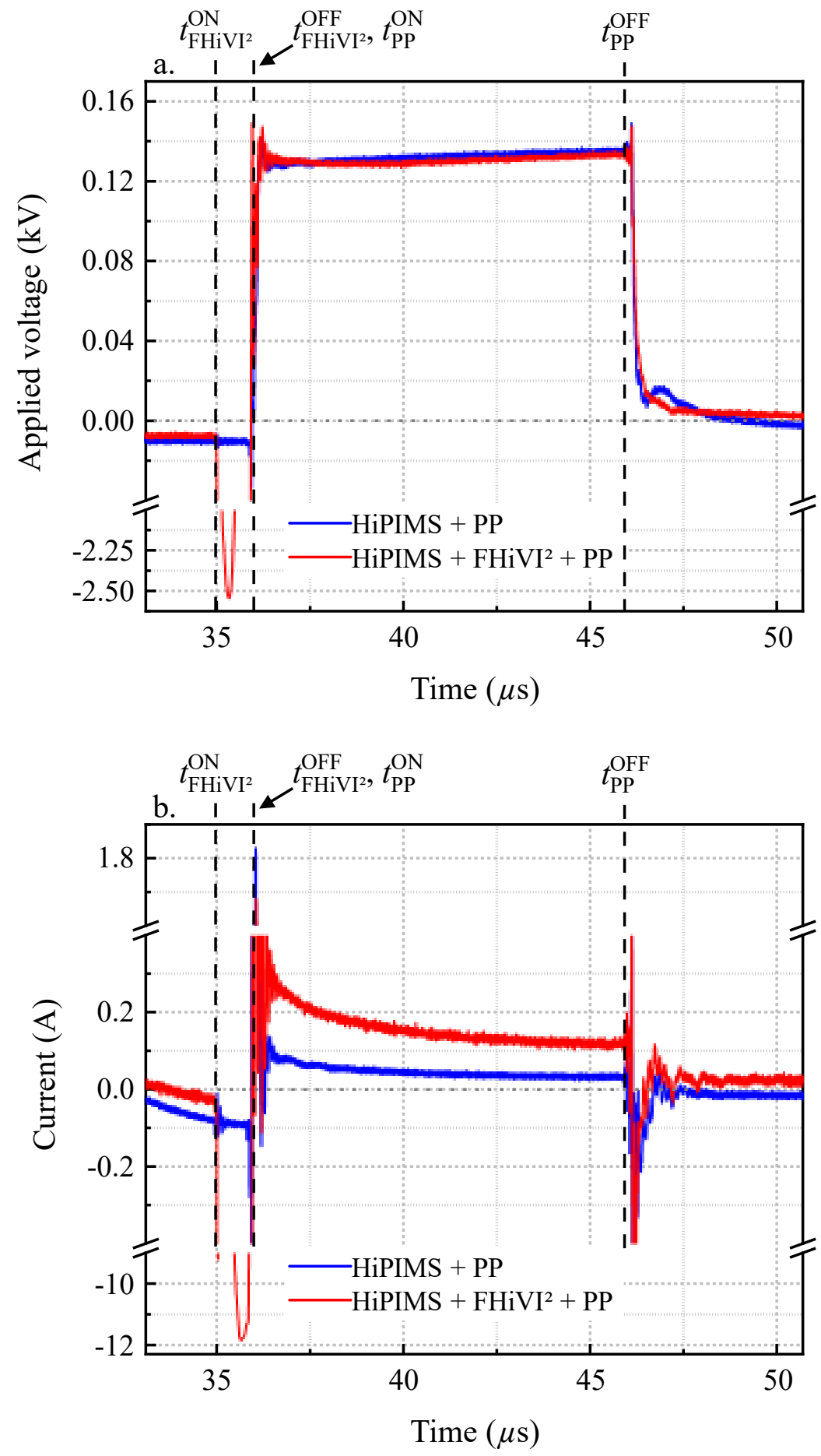

*Figure 6: Comparison of time evolution for (a) applied voltage and (b) measured current with (red) and without (blue) the FHiVI² regime. The axis breaks highlight variations during the PP regime while displaying the maximum values reached during FHiVI². Measurements were performed in pure argon at 1 Pa with $\Delta T_{FHiVI^2}$ = 35 $\mu$s and $V_{FHiVI^2}$ = -2.5 kV at $\nu$ = 200 Hz. The HiPIMS and PP regimes were operated under the following conditions: $T_{HiPIMS}$ = 20 $\mu$s, $V_{HiPIMS}$ = -0.4 kV, $T_{PP}$ = 10 $\mu$s, $V_{PP}$ = +150 V and $\Delta T_{PP} = t^{ON}_{FHiVI^2}$ + 1 $\mu$s.*

Thereby, Figure 6a and Figure 6b compare the time evolution of the applied voltage and the current measured with (red) and without (blue) the FHiVI² regime, respectively. Following FHiVI² application, Figure 6a shows that the applied voltage measured during the PP regime remains slightly unchanged (less than 3 V difference at the end of *III-b*). However, Figure 6b reveals an obvious increase in the measured current during time–domain *III-b*, by a factor of ~3.4 overall. The current $I_{PP}$ measurements confirm the enhanced charge production near the cathode induced by the FHiVI² regime. However, these measurements do not allow clear identification of the charges (ions, electrons) involved, particularly those originating from target sputtering (tungsten in this case). Moreover, knowledge of the ion–to–neutral ratio is crucial for assessing the effectiveness of the FHiVI² regime. A first approach to investigate this ratio, optical emission spectroscopy (OES) was first employed to evaluate the impact of the FHiVI² regime on the ionisation and radiative emission of W and $W^+$ species.

### 3.2. Optical emission spectroscopy: W II/W I ratio

To measure the contribution of the FHiVI² regime, time–resolved optical emission spectroscopy was carried out. The operating conditions of the HiPIMS, FHiVI² and PP regimes are, of course, identical to those described in the previous section.

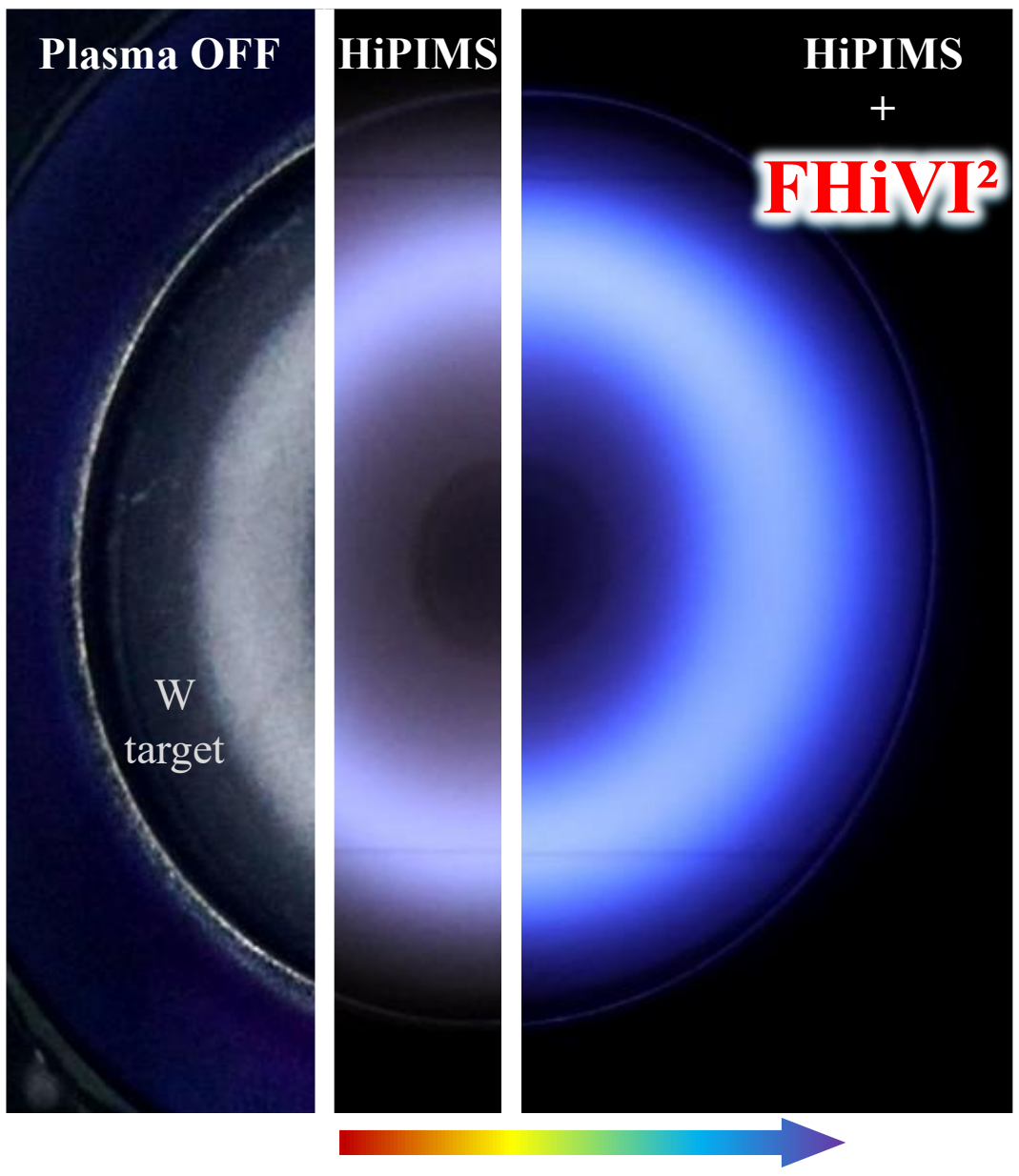

*Figure 7: Long-exposure front-view images of the target under three conditions, from left to right: plasma off, HiPIMS-only regime and with FHiVI² regime added.*

Figure 7 provides an overview of plasma emission under three different conditions. This broadband spectral comparison allows a first–order assessment of the adding effect of the FHiVI² regime to the pre–existing HiPIMS regime. Indeed, these images show a dominance of blue-shifted wavelengths when the FHiVI² regime is applied. This spectral shift therefore suggests, in first approximation, an increase in higher–energy excited species (Ar and W), consistent with the voltage applied during the FHiVI² regime. To gain further insight into the ionising effect of the FHiVI² regime, the following discussion considers the evolution of two emission lines from metallic species only, at 263.3 nm (W I: $E_i$ = 4.9 eV, $E_f$ = 0.2 eV) and 258.9 nm (W II: $E_i$ = 9.2 eV, $E_f$ = 4.4 eV, $E_I$ = 7.9 eV), corresponding to the de–excitation of W and $W^+$ species, respectively, according to NIST Atomic Spectra Database.

Figure 8a shows the time evolution of the $W^+$ emission line intensity in both cases, without (blue) and with (red) the application of the FHiVI² regime. Note that the case without FHiVI² corresponds to the combined HiPIMS + PP regime application, as described in Figure 4.

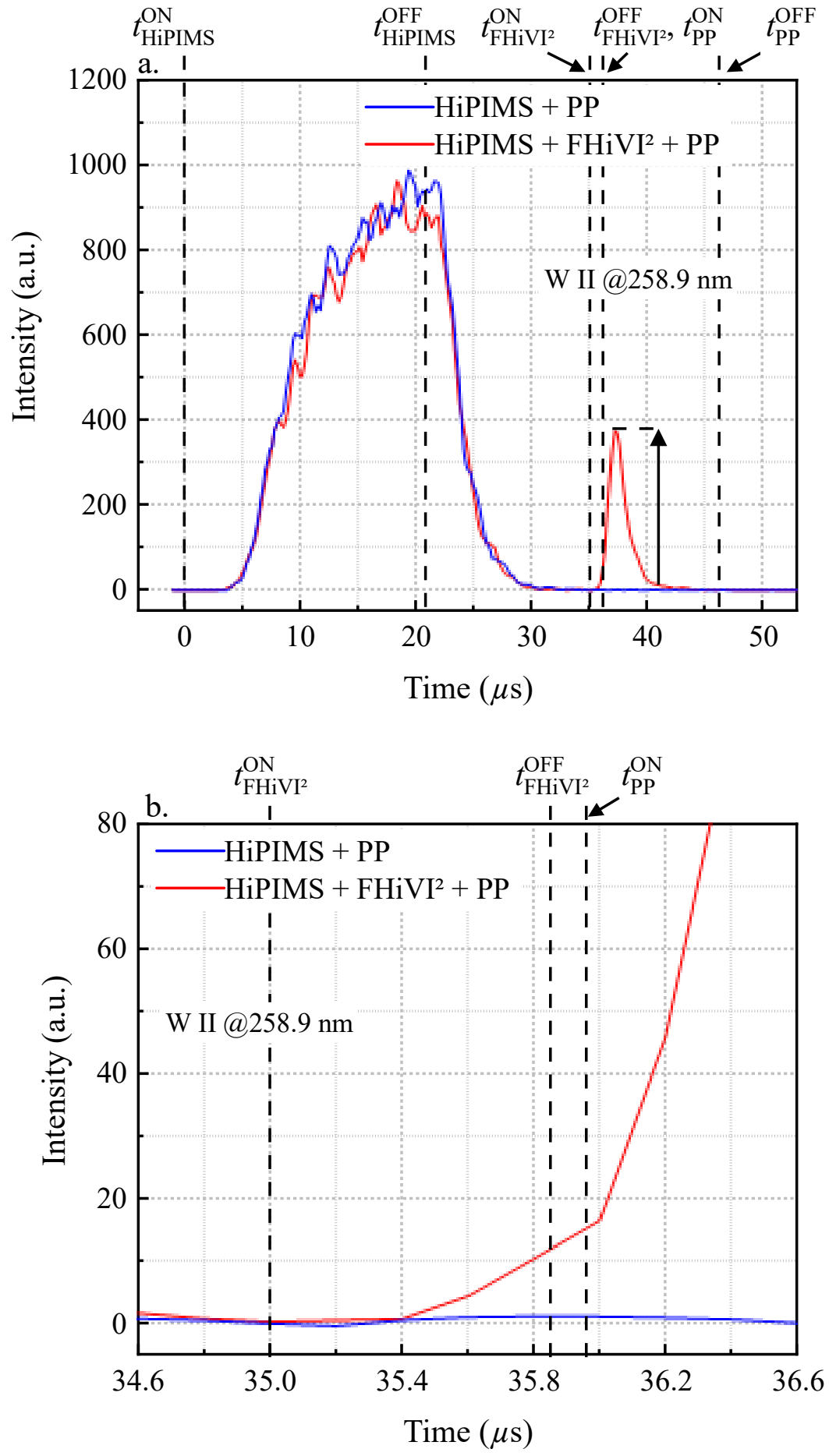

*Figure 8: (a) Time evolution of the W II emission line intensity at 258.9 nm, measured without (blue) and with (red) the application of the FHiVI² regime. To highlight the variations occurring during the FHiVI² regime, a zoom around its operating period is presented in (b). Measurements were performed in pure argon at 1 Pa with $\Delta T_{FHiVI^2}$ = 35 μs and $V_{FHiVI^2}$ = -2.5 kV at ν = 200 Hz. The HiPIMS and PP regimes were operated under the following conditions: $T_{HiPIMS}$ = 20 μs, $V_{HiPIMS}$ = -0.4 kV, $T_{PP}$ = 10 μs, $V_{PP}$ = +150 V and $\Delta T_{PP} = t^{ON}_{FHiVI^2}$ + 1 μs.*

In this first case, a single variation in the W II emission line intensity is measured from $t^{ON}_{HiPIMS}$ + 4 μs to $t^{ON}_{HiPIMS}$ + 30 μs. This variation is due solely to the HiPIMS regime, since the PP regime occurs much later in time. One may note that the latter does not appear to have any noticeable effect on these measured lines, since no intensity variation is observed over its duration. The observed delay between $t^{ON}_{HiPIMS}$ and the W II signal could be explained by the time required to produce the species required for $W^+$ ions excitation process. Indeed, the excitation of these ions results from several steps: sputtering of the W target following plasma ignition, ionisation of the sputtered W neutrals, and subsequent excitation of the resulting ions. The decay of the signal from $t^{OFF}_{HiPIMS}$ onwards reflects, amongst other, the decrease in the density of ionisable W neutrals as the plasma extinguishes, as suggested by the reduction in the V–I characteristics of the HiPIMS regime (*cf*. Figure 4).

Implementing the FHiVI² regime in the second case induces an additional temporal variation in the $W^+$ emission line intensity, observed after $t^{ON}_{FHiVI^2}$ and lasting approximately 5 µs. This duration, which exceeds that of the FHiVI² regime itself, suggests that the conditions leading to the ionisation and subsequent excitation of sputtered W neutrals remain favourable beyond $t^{OFF}_{FHiVI^2}$. Furthermore, as shown with Figure 8b, the intensity variation induced by the FHiVI² regime begins only after $t^{ON}_{FHiVI^2}$ + 0.4 µs, corresponding to the start of time–domain *II-b*, as shown with Figure 8b. Thus, this variation is concomitant with the sharp increase in the FHiVI² conduction current, whereas the variation induced by the HiPIMS regime begins later, relative to the HiPIMS conduction current, *i.e.* at $t^{ON}_{HiPIMS}$ + 4 µs. The difference in the time response observed for both regimes can be attributed to differences in their initial stages. The pre–existing population of species necessary for $W^+$ ion excitation (electrons, sputtered W neutrals, and $W^+$ ions) in the plasma volume, which were produced during the HiPIMS discharge, favour the earlier response of this line upon application of the FHiVI² pulse. This observation once again supports the discussion of the FHiVI² current, demonstrating that this regime operates in an environment already containing metallic species sputtered during HiPIMS regime to be ionised.

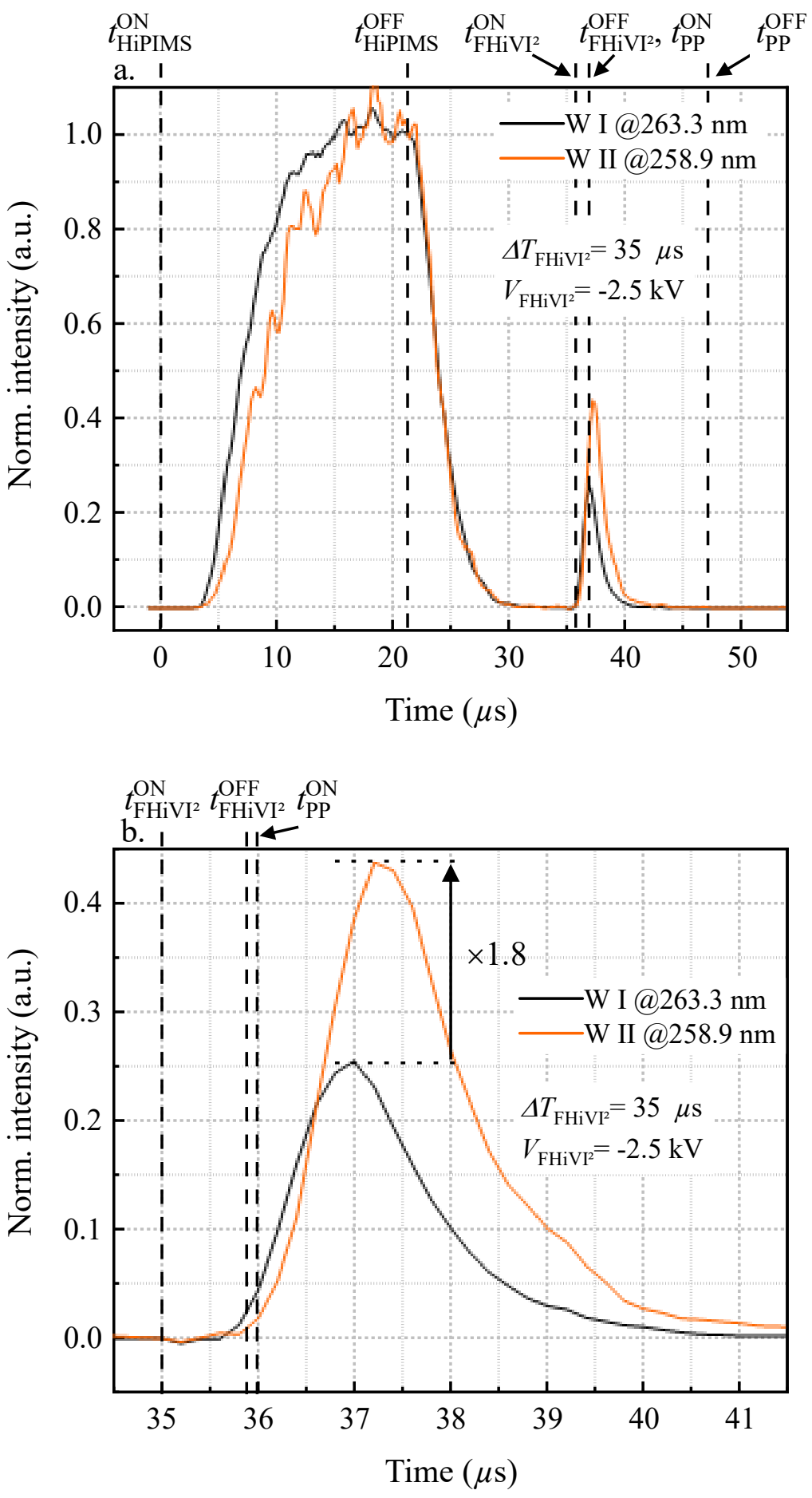

*Figure 9: (a) Time evolutions of the emission line intensities of W I (black) and W II (orange) at 263.3 nm and 258.9 nm, respectively, measured when the three regimes are applied as describe in section 3.1.3. To enable comparison between the lines, the W II emission line intensity is normalised to that of W I measured during the HiPIMS discharge. To highlight the variations induced by the FHiVI² regime, a zoom around its operating period is presented in (b). Measurements were performed in pure argon at 1 Pa with $\Delta T_{FHiVI^2}$ = 35 µs and $V_{FHiVI^2}$ = -2.5 kV at ν = 200 Hz. The HiPIMS and PP regimes were operated under the following conditions: $T_{HiPIMS}$ = 20 µs, $V_{HiPIMS}$ = -0.4 kV, $T_{PP}$ = 10 µs, $V_{PP}$ = +150 V and $\Delta T_{PP} = t^{ON}_{FHiVI^2}$ + 1 µs.*

To further evaluate the ionising potential of the FHiVI² regime, Figure 9 compares the emission lines of neutral W atoms (W I, black) and ions (W II, orange) under the condition where the three regimes—HiPIMS + FHiVI² + PP—are applied (*cf.* Figure 5). The intensities of the W I and W II emission lines are normalised using the HiPIMS regime as a reference.

The W I and W II intensity variations recorded during the HiPIMS regime are slightly delayed by 0.4 µs relative to one another, as observed in Figure 9a. As previously proposed, this delay is primarily due to the time required to produce enough excitable ions from sputtered neutral W atoms during the HiPIMS regime [34]. Furthermore, a nearly simultaneous decrease in the intensities of the W I and W II lines is observed upon plasma extinguishing at $t^{OFF}_{HiPIMS}$. This simultaneity demonstrates that the radiative mechanisms for both W and $W^+$ species at the considered wavelengths (263.3 nm and 258.9 nm, respectively) respond in the same manner.

In contrast, the application of the FHiVI² regime induces remarkable changes to the W I and W II emission time profiles. Firstly, and unlike the HiPIMS regime, the W I and W II intensity signals rise at the same time, starting from $t^{ON}_{FHiVI^2}$ + 0.4 µs, as observed in Figure 9b. This confirms that W and $W^+$ species are already present in the volume due to the HiPIMS regime and persist at least until $t^{ON}_{FHiVI^2}$ + 35 µs. In other words, a proportion of neutral W atoms and ions remain in the HiPIMS post–discharge and contribute to the radiative emission induced by the FHiVI² regime, at least during these initial moments. Furthermore, the peak intensity ratio of the W II line relative to W I is higher by a factor of approximately 1.8. To explain this increase, two hypotheses can be proposed: (i) the excitation and de–excitation mechanisms of pre–existing $W^+$ ions in the volume are enhanced, or (ii) the $W^+$ ions volume density increases through ionisation of the already present neutrals, thereby relatively decreasing the neutral W density. The second hypothesis is favoured in light of the previously stated observations, namely the near–simultaneous response of the W I and W II lines at $t^{OFF}_{HiPIMS}$ and the intensity increase in both lines nearly concomitant with the rise in FHiVI² conduction current at $t^{ON}_{FHiVI^2}$ + 0.4 µs (*cf.* time–domain *II-b*). Furthermore, hypothesis (ii) is also supported by the 0.4 µs delay observed between the maxima of the W II and W I lines. This latter result, combined with the overall duration of approximately 5 µs for the W II emission, indicates on one hand that ionisation continues well beyond $t^{OFF}_{FHiVI^2}$ compared to $t^{OFF}_{HiPIMS}$, and on the other hand reveals the time required for collisional electron cooling after $t^{OFF}_{FHiVI^2}$.

The differences induced by the FHiVI² regime compared to the HiPIMS regime involve substantial modifications to the proportion of W and $W^+$ species. To quantify this change, the $Q^{OES}$ ratio is calculated from the integrals under the W I and W II curves, as expressed in Equation 1.

$$Q^{OES} = \frac{[W\ II/W\ I]_{Electrical\ regime}}{[W\ II/W\ I]_{HiPIMS+PP}} \qquad \text{Equation 1}$$

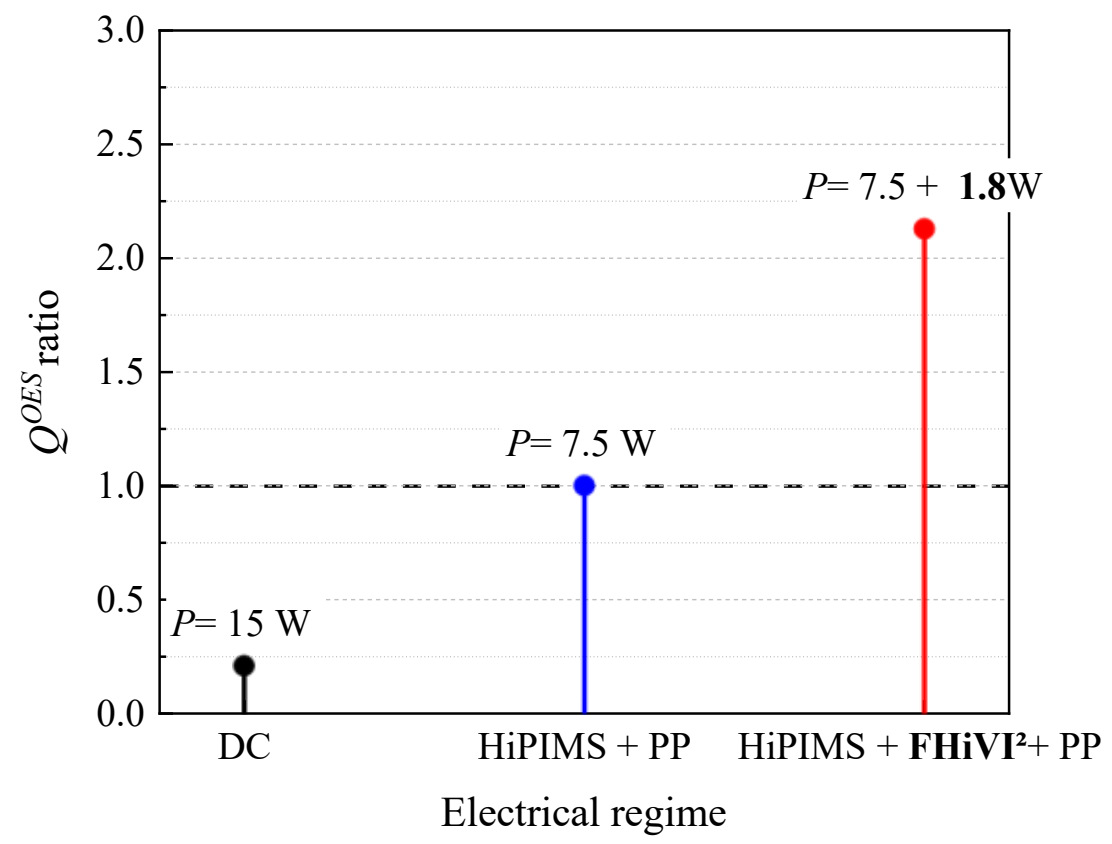

*Figure 10: Comparison of the $Q^{OES}$ ratios computed from the integrals of the W II and W I lines, normalised to that calculated for the HiPIMS + PP regime, for different operating regimes: DC, HiPIMS + PP without FHiVI², and HiPIMS + FHiVI² + PP.*

Plotted in Figure 10, the $Q^{OES}$ ratio is compared across three discharge regimes: DC at 15 W, HiPIMS + PP without and with FHiVI² regime. Firstly, Figure 10 shows that the $Q^{OES}$ ratio for the DC regime is lower than that for HiPIMS + PP regime, although the mean power delivered during the DC regime is twice that of HiPIMS. Obviously, this result is expected given prior studies demonstrating that the HiPIMS regime enhances the ionisation rate of sputtered species. Thus, adding the FHiVI² regime to HiPIMS + PP increases the $Q^{OES}$ ratio by a factor of approximately 2.2. This increase is even more significant given that the FHiVI² regime represents only 20% of the mean power delivered by the HiPIMS regime. In this application example, this result thus highlights the absolute need to consider the shape of the voltage applied to the cathode to optimise the ionisation processes of sputtered metal species.

In summary, the various results from the radiative emission of W and $W^+$ species reveal numerous modifications following the application of the FHiVI² regime, particularly an increase in the W II/W I line intensity ratio ($Q^{OES}$). However, optical emission spectroscopy (OES) provides only partial and qualitative information on the actual variation in $W^+$ ion density, since the excitation and de–excitation processes also depend on electron density, reaction coefficients, etc. To complement these analyses, mass spectrometry (MS) diagnostics were therefore deployed to provide information on the temporal evolution of $W^+$ ion fluxes.

### 3.1. Mass spectrometry: $W^+$ ion fluxes

To gain further insight into the interaction of the FHiVI² regime with the other regimes, as well as its ionising effectiveness on the metallic species of interest, time– and energy–resolved mass spectrometry was carried out. To minimise perturbations in the mass spectrometry measurements, the maximum voltage amplitude applied during the FHiVI² $V_{FHiVI^2}$ regime was limited to 2 kV.

Figure 11 reports the time– and energy–resolved distribution of $W^+$ ions under conditions without (a) and with

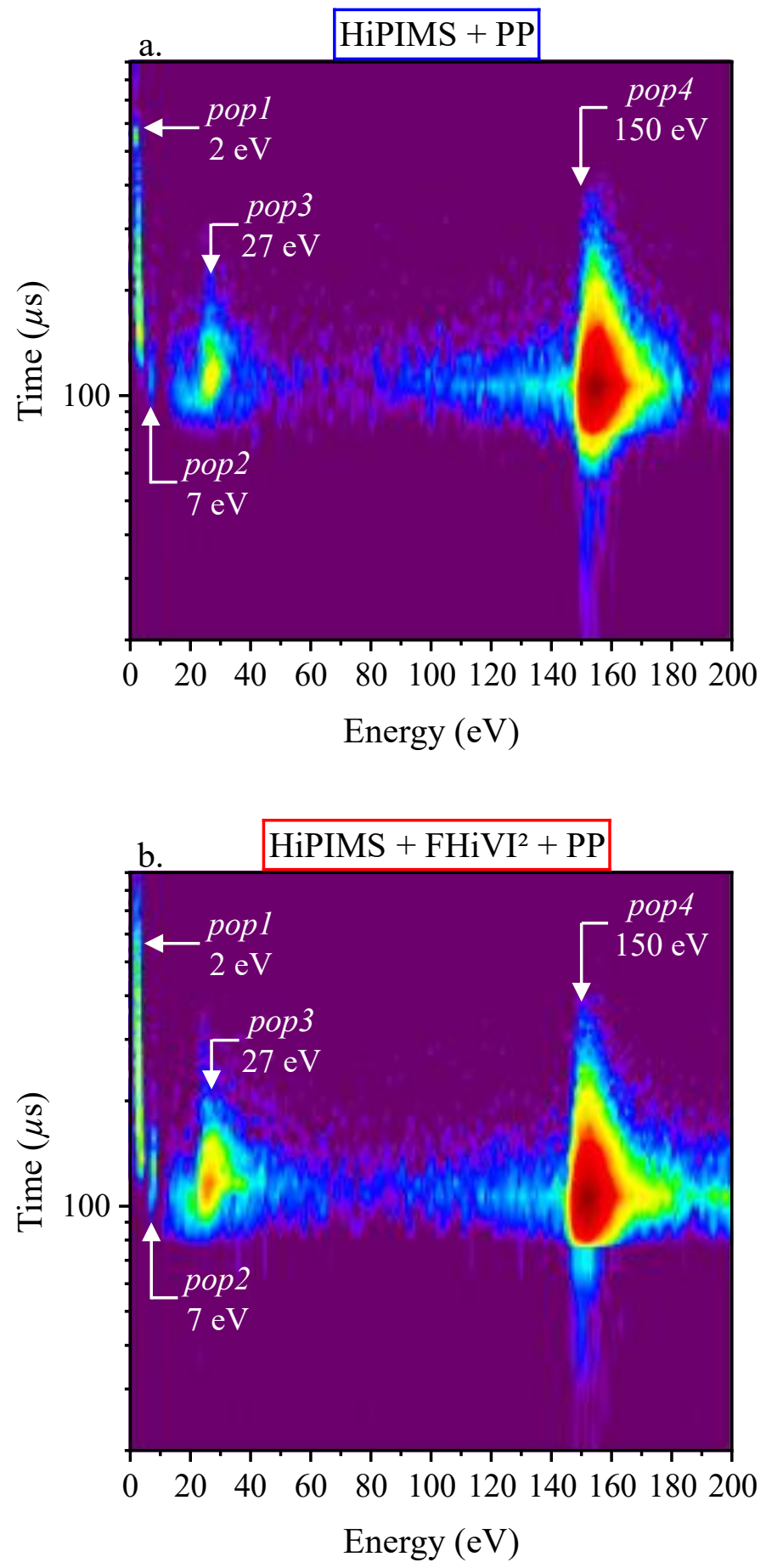

*Figure 11: Two–dimensional, time–resolved representations of the energy distributions of $W^+$ ions for two cases: (a) HiPIMS + PP and (b) HiPIMS + FHiVI² + PP. Measurements were performed in pure argon at 1 Pa with $\Delta T_{FHiVI^2}$ = 35 µs and $V_{FHiVI^2}$ = -2 kV at ν = 200 Hz. The HiPIMS and PP regimes were operated under the following conditions: $T_{HiPIMS}$ = 20 µs, $V_{HiPIMS}$ = -0.4 kV, $T_{PP}$ = 10 µs, $V_{PP}$ = +150 V and $\Delta T_{PP} = t^{ON}_{FHiVI^2}$ + 1 µs.*

(b) the FHiVI² regime applied, together with the HiPIMS and PP regimes. As stated in the experimental setup section, the timescale is associated to the time at which the $W^+$ ions enter the apparatus upon $t^{ON}_{HiPIMS}$. However, these ion arrivals result from a combination of processes spanning ion production (in both space and time) and transport, which are strongly influenced by the timing and duration of the different electrical regimes. None of are resolved here.

From Figure 11, observed variations can be categorised into four distinct energy populations, regardless of whether case (a) or (b) is considered. To identify these populations, the following discussion draws on the work reported by A. el Farsy *et al.* [35, 36], since the corresponding distributions exhibit similar features even in a different study system. This short discussion only provides an overview of the results as the aim is to discuss the effectiveness of the FHiVI² regime.

Designated *pop1*, the first variation is observed between 1 to 5 eV and exhibits a profile consistent with a Gaussian distribution. This population may correspond to thermalised $W^+$ ions with a quasi–isotropic velocity distribution and diffusive transport. Thus, *pop1* could arise both from the ionisation of thermalised neutral W atoms and from the collisional thermalisation of higher–energy $W^+$ ions populations within the volume.

The second population, *pop2*, is observed between 6 to 10 eV and likewise exhibits a Gaussian–like profile. This population may arise from incomplete collisional thermalisation of higher–energy $W^+$, leaving ions with a mean energy still greater than that of *pop1*. Accordingly, *pop2* is identified as a population of quasi–thermalised ions, characterised by an anisotropic velocity distribution and quasi–diffusive transport.

By contrast, the third variation is observed over a broader energy range from 20 to 50 eV, and displays a non-Gaussian distribution. More precisely, this profile is characterised by a steep rise to a maximum value at 27 eV followed by a slow decay (Thompson–Stepanova–like profile [37, 38]). Consequently, this third population, *pop3*, is considered non thermalised with an anisotropic velocity distribution and essentially ballistic transport. To explain such a profile and determine its origin, *pop3* has been extensively investigated in the literature under various hypotheses: ionisation of high–energy neutral sputtered atoms [39], reflection of fast ions on the target [40] and the influence of *spokes* [41–44], an azimuthal structure observed in the $\mathbf{E} \times \mathbf{B}$ plane [45, 46]. Taken together, these hypotheses support the interpretation that this population is representative of ions generated in the plasma directly after target sputtering. As a result, *pop3* would chronologically be the first population present in the plasma volume and would likely constitute the source of the others *pop1*, *pop2*. Thus, *pop3* appears to provide the clearest insight into the ionisation of sputtered neutral species.

Finally, the fourth population, *pop4*, is observed over a much higher energy range, approximately 100 to 185 eV, with an asymmetric profile. This population is strongly linked to PP regime, which is why we refer to it as a population of repelled ions. Given the characteristics of the PP regime, this population is likely fed by the acceleration of $W^+$ ions near the cathode, particularly those initially present in the cathode sheath, which would otherwise have been lost through thermalisation, or loss to the cathode [47].

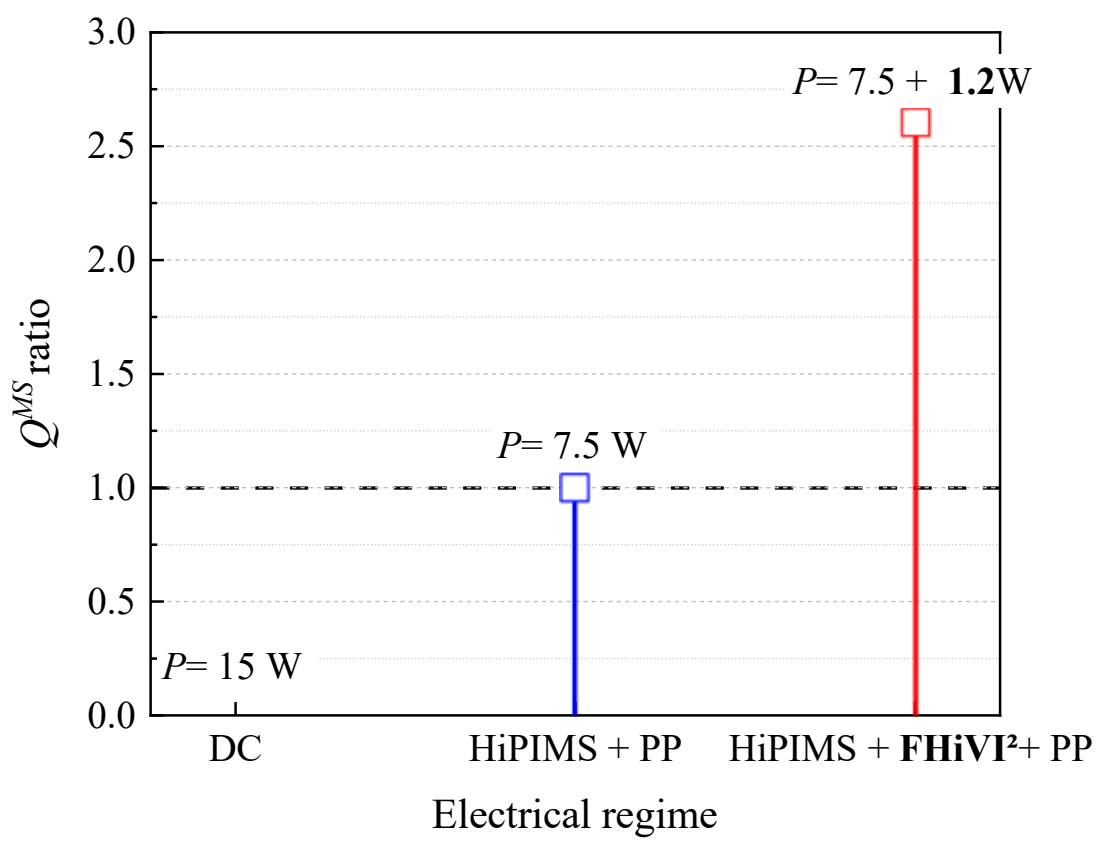

*Figure 12: Comparison of the $Q^{MS}$ ratios computed from the time– and energy–integrated signal of pop3, normalised to that calculated for the HiPIMS + PP regime, for HiPIMS + PP without FHiVI², and HiPIMS + FHiVI² + PP.*

Consequently, *pop4* is necessarily enriched by all previously identified lower–energy populations. However, these measurements suggest that *pop4* is largely sourced from *pop3*: the two populations exhibit very similar energy-distribution profiles, and the peak energy of *pop4* (~ 153 eV) is consistent with the peak energy of *pop3* (~ 27 eV) augmented by the energy imparted to ions during acceleration in the PP regime (by theoretically $V_{PP}$ = +125 V).

A comparison of the measurements in Figure 11 already indicates that the FHiVI² regime produces a significant increase in intensity relative to *pop3*. To better quantify the contribution of the FHiVI² regime to the $W^+$ ion flux relative to *pop3*, the signal was integrated over time and energy. The resulting value was then normalised on the HiPIMS + PP regime and referred $Q^{MS}$ ratio.

Figure 12 compiles the $Q^{MS}$ results only for two conditions: HiPIMS + PP without (blue) and with (red) the FHiVI² regime. In fact, measurements in the DC regime, performed with the mass–spectrometer settings kept strictly identical, do not yield any signal that exceeds the noise background. This further highlights the low $W^+$ ion flux generated compared with the other regimes, even though the DC regime delivers a higher mean power than the HiPIMS regime here. More importantly, adding the FHiVI² regime to HiPIMS + PP significantly increases the $W^+$ ion flux, by a factor $Q^{MS} \approx 2.6$. As mentioned with OES analysis, this result is even more striking given the relative mean power supplied by the two regimes, since FHiVI² accounts for only 14% of the total mean power. On the other hand, the differences between the results obtained by MS and those derived from the OES measurements discussed above are not solely due to the decrease in set $V_{FHiVI^2}$ for MS measurements since the latter represents a loss of only 20%. Rather, they arise from the nature of the diagnostics themselves. Indeed, the OES is a non–local diagnostic that provides information only on the species that were excited, which limits the semi-quantitative nature of the estimate. By contrast, MS diagnostic measures the local ion flux that passing through the apparatus orifice. Therefore, these results are strongly dependent on the orifice position relative to the cathode, owing to the angular dependence of species transport.

## 4. Conclusion

This article investigates a novel pulsed regime named FHiVI², an electrical regime mainly intended to improve the ionisation of species previously sputtered from another sputtering source, here we choose HiPIMS. Amongst other benefits, the purpose of this regime is to provide a device that is easy to implement and capable of enhancing the ionisation rate of metal vapour from a pre-existing sputtering regime, particularly the HiPIMS regime here. Thus, this article aims to demonstrate the feasibility and relevance of such a regime using diagnostics to assess the ionisation rate achieved with FHiVI².

This regime is characterised by a relatively short application duration (sub-microsecond) at applied voltages exceeding typical HiPIMS voltages and implying fast voltage variations. These variations serve to re-accelerate residual electrons present during the HiPIMS post-discharge, resulting in a rapid increase in the FHiVI² current measured in time–domain *II-a*. This re-acceleration causes the ionisation of species in the volume, leading to the growth of the FHiVI² current observed in time–domain *II-b*. Moreover, measurements of the PP current during time–domain *III-b* corroborate this. Consequently, the application of the FHiVI² regime implies a significant modification, at least in the target vicinity.

Furthermore, optical emission spectroscopy and the mass spectrometry both indicated significant enhancement of the ionisation rate due to the FHiVI² regime. More specifically, mass spectrometry measurements reveal a 2.6-fold increase in the $W^+$ ion flux relative to HiPIMS, whereas the FHiVI² regime increases the total mean power by only approximately 1.1-fold. These initial results mark the early development of the FHiVI² regime. Given the highly encouraging outcomes, further research on this innovative regime must be pursued.

**Acknowledgments**

This work has been funded by the French Agence Nationale de la Recherche (ANR) under project SYNERGY (ANR-20-CE05-0013).

The authors are grateful to Bruno MAZIERE, power electronic expert, for insightful discussions on high-power circuitry management.